\documentclass[sigconf]{acmart}

\setcopyright{acmcopyright}
\copyrightyear{2018}
\acmYear{2018}
\acmDOI{XXXXXXX.XXXXXXX}

\acmConference%
    [EuroPLoP 2022]%
    {25th European Conference on Pattern Languages of Programs (EuroPLoP 2022)}%
    {6.-10.~July 2022}%
    {Kloster Irsee, Germany}
    
\acmPrice{15.XX}
\acmISBN{978-1-4503-XXXX-X/18/06}

\usepackage[inline]{enumitem}
\usepackage{graphicx}
\usepackage{textcomp}
\usepackage{tikz}
  \usetikzlibrary{shapes,arrows,automata,shadows,positioning,arrows.meta}
\usepackage{multicol}
\usepackage{multirow}
\usepackage{newfloat}
\usepackage{placeins}

\usepackage{selnolig}  
\usepackage{soul}
\usepackage{upquote}

\usepackage{matlab-prettifier}
\usepackage{tcolorbox}
\tcbuselibrary{listings}
\newtcblisting[auto counter]{tlisting}[3][]{
    left = .8em,
    label=#3,
    listing only,
    listing options={
        style=Matlab-editor,
        basicstyle=\footnotesize\ttfamily,
        breaklines         = false,
        escapeinside={<@}{@>},
    },
    fonttitle=\bfseries,
    title=Listing \thetcbcounter: {#2},
    #1}
\definecolor{remark}{RGB}{034,139,034}
\definecolor{keyword}{RGB}{180,80,0}
\definecolor{KEYWORD}{RGB}{180,80,0}
\definecolor{stringcolor}{RGB}{160,032,240}
\definecolor{STRINGCOLOR}{RGB}{160,032,240}
\definecolor{persistentcolor}{RGB}{0,163,137}
\definecolor{commentcolor}{RGB}{034,139,034}
\definecolor{ecommentcolor}{RGB}{100,100,100}
\newcommand{\kw}[1]{{\color{keyword}#1}}
\newcommand{\mstring}[1]{{\color{stringcolor}#1}}
\newcommand{\qstring}[1]{{\color{stringcolor}\textquotesingle#1\textquotesingle}}
\newcommand{\mcomment}[1]{{\color{commentcolor}#1}}
\newcommand{\ecomment}[1]{{\color{ecommentcolor}#1}}

\newcommand\Tstrut{\rule{0pt}{2.6ex}}         
\newcommand\Bstrut{\rule[-0.9ex]{0pt}{0pt}}   

\newdimen\hfuzz  

\begin{document}

\title[Injection testing]{Injection testing backed refactoring}

\author{Thomas Mejstrik}
\affiliation{%
    \institution{University of Vienna}
    \city{Vienna}
    \country{Austria}
}
\email{thomas.mejstrik@gmx.at}
\orcid{0000-0003-2801-08282}

\author{Clara Hollomey}
\affiliation{%
    \institution{Austrian Academy of Sciences}
    \city{Vienna}
    \country{Austria}
}
\email{clara.hollomey@oeaw.ac.at}

\begin{abstract}

Injection-based testing while refactoring is a pattern that 
minimizes the need for manual editing 
when altering the behaviour of a code base. 
Neither does it rely on a compilation or a linking process 
nor does it make assumptions on the structure of the code. 
Thus, it can be particularly useful for refactoring code 
that has been written in scripting languages, 
and specifically targets the research and engineering context. 
We describe the pattern and propose a set of functions for its application. 
The applicability of code injection for refactoring 
is highlighted via specific examples for deriving unit and integration tests. 
Finally, we comment on the customizing of the pattern 
and give practical advice for its implementation.


    
    
\end{abstract}




\maketitle

\section{Introduction}
We introduce a pattern for injecting test code to a code base 
to facilitate its \emph{refactoring}, 
i.\,e.~changing the internal structure of code 
without affecting its external behaviour~\cite{Opd1992}.
The injection testing pattern enlarges the set of code bases 
to which testing provisions can be added retrospectively.
The pattern allows inferring the internal behaviour of code 
by defining designated interception points as comments in the source code. 
No strong requirements are placed on the programming style of the code,
which is especially important for refactoring legacy code.

\begin{figure}[tb]
    \centering
    \Description{%
        Refactoring cycles
        }

{
\def \balltextwidth {1.6cm}
\def \radius {1.4cm}
\def \lmargin {25}
\def \rmargin {23}
\def \shift {0}

\tikzstyle{roundrect}=[rectangle, rounded corners, draw=black, thick, 
           inner sep = 2pt,
           minimum size = 1.6cm,
           text width = 1.6cm,
           align=center,
           fill = black!10,
           ]
\begin{tikzpicture}
\node[roundrect]
at ({180+\shift}:\radius) 
    {\parbox[t]{\balltextwidth}{%
        \begin{center}\emph{Write a  succeeding test}\end{center}%
    }};

\node[roundrect]
at ({\shift}:\radius) 
    {\parbox[t]{\balltextwidth}{%
        \begin{center}\emph{Refactor}\end{center}%
    }};

\draw[-{Latex[length=2mm]}] ({180+\shift+\lmargin+10}:\radius) 
arc ({180+\shift+\lmargin+10}:{360+\shift-\rmargin}:\radius);
    
\draw[-{Latex[length=2mm]}] ({0+\shift+\lmargin+10}:\radius) 
arc ({0+\shift+\lmargin+10}:{180+\shift-\rmargin-6}:\radius);
\end{tikzpicture}%
}    
    \caption{Refactoring cycles: 
    (1) In refactoring, one often has to start with the 
    \emph{Refactor} step.
    (2) Using the injection testing pattern one can start with the 
    \emph{Write a succeeding test} step, 
    and thus, make the refactoring step safe.\newline
    \hspace*{1em}
    The refactoring step
    not only concerns the production code,
    but also the accompanying test suite.
    }
    \label{fig_refactorcycle}
\end{figure}
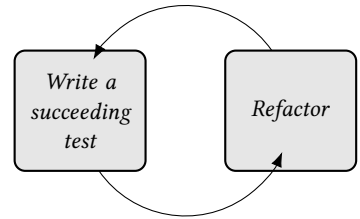

Techniques for refactoring are well established, see e.\,g.~\cite{Fow2018,KMZ2020,Dall2012,BMS2014,AACCG2017,NV2017,KR2011,HRS2013},
and often rely on the existence of
an accompanying test suite~\cite{Fea2004}, 
as depicted in Figure~\ref{fig_refactorcycle}.

The idea is to start from a succeeding test suite, 
apply the desired changes, and confirm that the observable behaviour did not change
by running the test suite again.

For code without testing provisions, however, 
this classical refactoring cycle indicates a chicken-and-egg problem: 
It is unclear how to write a test suite
prior to touching the code
when no such test suite is already present.
Consequently, in the absence of tests
there is always the risk of unwittingly changing the behaviour of the code 
and introducing new bugs. 
Thus, the challenge is intercepting a code`s internal behaviour 
while at the same time minimizing the amount of manual editing.

\subsection{Existing solutions}
\label{sec_existing_solutions}

Common interception points arise from the compiler toolchain and have been summarized under the term ``seams'', \emph{
``%
\dots a place where you can alter behaviour in your program without editing in
that place.%
''
}~\cite[Chapter~4]{Fea2004}.

\emph{Pre-processing seams} can be accessed via a text replacement engine 
that automatically changes the source code just before interpretation or compilation.
Besides the need for maintaining such an engine in languages that do not feature a pre-processor, 
the usage of such seams necessitates a new compilation/interpretation cycle 
for every change that is applied to the code.

\emph{Link seams} can be used in languages where the compiler/interpreter 
produces an intermediate representation of the code. 
Intercepting calls to other intermediate representations 
allows for the execution of arbitrary code. 
Similarly to the pre-processor approach, 
an additional linking step is required after each refactoring iteration, 
impeding the practical applicability of the approach.

\emph{Object seams} are not compilation-related and 
use overloading and polymorphism for intercepting calls to class methods. 
This requires that both, the programming language, 
and the programming style of the code at hand, 
support object orientation and dependency injection at least to some degree. 

All of the above approaches pose requirements 
on either the programming language or the programming style 
in which the code has been written. 
Not every code meets those requirements.


%

\subsection{Our motivation}
\label{sec_motivation}

Scripting languages are commonly used for numerical computations in 
engineering, scientific, and educational contexts. 
Also in those contexts, programs are often written incrementally
and by frequently changing authors.
This can lead to large code bases with insufficient 
testing provisions~\cite{BK2014}.

In fact, an informal survey of software papers published 
between 2015 and 2021 for Matlab, 
one of the most common programming languages in academia~\cite{mw_matlabusage}, 
indicates that only a small fraction of submissions includes tests,
let alone automated tests using some specialized testing framework,
as indicated in Table~\ref{tab_ut_usage}.
This makes it hard to
verify the correctness of the published results.

Still, scientific code often needs to be refactored,
and when such code has been written in a scripting language,
the absence of a pre-processor, 
a compilation and a linking step limits the options for intercepting 
the internal workings of the code. 

\begin{table}[tb]
{
\centering
\begin{tabular}{cccc}
    \multirow{2}{*}{Journal$^\dagger$} & Matlab\Tstrut &        unit         &     automatic      \\
                                       &  submissions  &        tests        &     unit tests     \\ \hline
            \emph{JOSS}\Tstrut         &      25       & \phantom{0}9 (36\%) &      3 (12\%)      \\
             \emph{ACM TOMS}           &      19       & \phantom{0}9 (47\%) &      2 (11\%)      \\
               \emph{JORS}             &      22       & \phantom{0}8 (36\%) & 2 \phantom{0}(9\%) \\
         \emph{J.\ Stat.\ Softw.}      &      18       & \phantom{0}8 (44\%) & 1 \phantom{0}(6\%) \\
             \emph{Softw.\ X}          &      78       &      11 (14\%)      & 5 \phantom{0}(6\%) \\
        \emph{Softw.\ Imp.}\Bstrut     &      13       & \phantom{0}2 (16\%) & 0 \phantom{0}(0\%)
\end{tabular}
}

\caption[Number of Matlab submissions with unit tests]%
{Number of publications of free Matlab software whose source code is still available,
in selected scientific journals between 2015 and 2021, 
contrasted with the number of submissions additionally including unit tested code 
and automatic unit testing provisions.
\newline
\normalfont
\footnotesize{$^\dagger$%
    \emph{JOSS}: Journal of Open Source Software, 
    \emph{ACM TOMS}: ACM Transactions on Mathematical Software,
    \emph{JORS}: Journal of Open Research Software, 
    \emph{J.~Stat.~Softw.}: Journal of Statistical Software,
    \emph{Softw.~X}: Software X,
    \emph{Softw.~Imp.}: Software Impacts.
    }
}
\label{tab_ut_usage}
\end{table}

\subsection{Overview}
In Section~\ref{sec_inject},
we present our injection testing pattern 
and propose a set of functions for its efficient implementation

We show how the pattern can be applied to write \emph{unit tests} in Section~\ref{sec_inject_unit}. 
The  code injection pattern is further exemplified in Section~\ref{sec_inject_integration},
where we discuss how the injection testing pattern
can be used to write \emph{integration tests}.\footnote{
\emph{Unit tests} test well-defined, self contained parts of code, whereas
\emph{integration tests} test the code together with some of its dependencies with respect to specified functional requirements.}

Finally, in Sections \ref{sec_implementation} and~\ref{sec_ttest}
we discuss some points regarding the implementation of our pattern,
including the presentation of the Matlab/Octave\footnote{%
\emph{Octave} is a free implementation of the Matlab language.} 
unit test framework~\emph{TTEST}.

\section{Injection testing pattern}
\label{sec_inject}

\subsection{Context}
Given a function to which tests shall be added,
in the following also referred to as \emph{system under test},
that has no or insufficient testing provisions,
this pattern can be applied.

\subsection{Problem}
Refactoring without having tests 
in place bears the risk of breaking the code. 
Thus, the challenge is to add tests to existing code 
while at the same time limiting the impact of any required manual editing.


\subsection{Forces}
In situations where the requirements on the code and programming language outlined in section~\ref{sec_existing_solutions} are not met, it can be hard to infer the internal workings of code without a considerable amount of manual editing.
The manual editing of code, however, always bears the risk of unwittingly altering its behaviour. Without prior tests in place, it is hard to detect such unintended changes.

\subsection{Solution}
\label{sec_solution}
We add tests by injecting\footnote{%
    We use the term \emph{code injection} instead of 
    \emph{code instrumentation},
    to stress the point that we do not just add logging or similar functionality, 
    but arbitrary code to an existing code base.}
test code, executed at runtime, defining specific entry and exit points,
and thus minimize the amount of code changes.
To minimize the risk that the injected code alters the behaviour
of the code base in an unintentional way we, first, propose the following pattern:

\begin{center}
\large
\emph{Inject arrange - Act - Assert},
\end{center}
optionally supplemented by an \emph{Inject setup} stage at the very beginning and a
\emph{Tear down} stage at the very end.

\begin{figure}[tb]
    \centering

\usetikzlibrary{positioning}
\usetikzlibrary{shapes.geometric, arrows}
{\small
\tikzstyle{splitrect}=[rectangle, rounded corners, draw=black, thick, 
           anchor=north west,
           text width = 6.1cm,
           rectangle split, 
           rectangle split parts = 2,
           inner sep = 4pt,
           align = center,
           fill = black!10,
           ]
           
\tikzstyle{arrow} = [thick,-{Latex[length=2.5mm]},>=stealth]

\begin{tikzpicture}[->,>=stealth',node distance=0.5cm]

\node[ splitrect
     ] (SETUP) {
    \bfseries{Inject Setup}
    \nodepart{second}
    \vbox{\begin{itemize}[leftmargin=*]
    \item Inject jump to the section of interest via~\texttt{\kw{gotoat}}
    \item Inject retrieval of results after the section of interest via~\texttt{\kw{captureat}}
    \item Inject return after the section of interest via~\texttt{\kw{returnat}}
    \end{itemize}}
};

\node[ splitrect,
       below =of SETUP,
     ] (ARRANGE) {
    \textbf{Inject Arrange}
    \nodepart{second}    
    \vbox{\begin{itemize}[leftmargin=*]
        \item Inject code to set the workspace via \texttt{\kw{assignat}}
    \end{itemize}}
}; 

\node[ splitrect,
       below =of ARRANGE,
       text width = 4cm,
     ] (ACT) {
    \textbf{Act}
    \nodepart{second}    
    \vbox{\begin{itemize}[leftmargin=*]
        \item Start the system under test
    \end{itemize}}
}; 

\node[ splitrect,
       below =of ACT,
     ] (ASSERT) {
    \textbf{Assert}
    \nodepart{second}
    \vbox{\begin{itemize}[leftmargin=*]
    \item Retrieve the results via \texttt{\kw{captureat}}
    \item Check the results (using an assertion framework)
    \end{itemize}}
};

\node[ splitrect,
       below =of ASSERT,
     ] (TEARDOWN) {
    \textbf{Tear down}
    \nodepart{second}
    \vbox{\begin{itemize}[leftmargin=*]
    \item Remove all code injections via \texttt{\kw{clearat}}
    \end{itemize}}
}; 

 \path[arrow]
 (SETUP)      edge     (ARRANGE)
 (ARRANGE)    edge     (ACT)
 (ACT)        edge     (ASSERT)
 (ASSERT)     edge     (TEARDOWN)
 ;
 
 \draw [arrow] ([shift={(25mm,0mm)}]ASSERT.north) -- ([shift={(25mm,0mm)}]ARRANGE.south); 

\end{tikzpicture}
}
    \caption{Injection testing pattern: 
    After an optional setup, in the \textit{Inject Arrange} stage, 
    the code to set the desired workspace is injected. 
    After the execution of the system under test in the \textit{Act} stage, 
    the results are retrieved and checked in the \texttt{Assert} stage. 
    The final tear down concludes the pattern.}
    \label{fig_pattern}
    \Description{Injection testing pattern}
\end{figure}
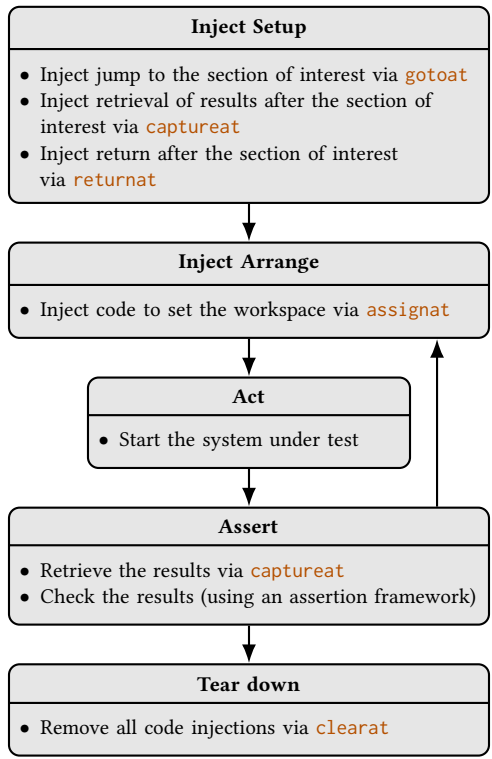

Second, we propose the following set of key functions:
\begin{itemize}
\item \texttt{\kw{gotoat}}:    Jumps to an arbitrary line of code
\item \texttt{\kw{assignat}}:  Assigns values to variables
\item \texttt{\kw{captureat}}: Stores the current state for later use
\item \texttt{\kw{returnat}}:  Returns from the function
\item \texttt{\kw{clearat}}:   Removes all injected code
\end{itemize}
The correspondence between the key functions and 
the stages of the injection testing pattern are depicted in Figure~\ref{fig_pattern}.

The injection testing pattern closely follows that of a classical unit test~\cite[Chapter~3]{Khor2020}.
In the \emph{Arrange} section, objects are initialized and data is passed to the system under test.
In the \emph{Act} section, the system under test is invoked with the arranged data.
In the \emph{Assert} section, the actions and results of the system under test 
is compared with the expected actions and results.

One key difference to the unit test pattern is that the \emph{Inject arrange} section 
is in fact not executed before the system under test is invoked.
This section only arranges the code which shall be executed when the system under test is running.%

It would be possible to also move the \emph{Assert} section 
into the system under test.
However, this would couple the unit tests and the system under test more strongly and 
increase the amount of injected code into the system under test.
Both aspects may increase the chance of unintentionally changing the behaviour of the system under test at runtime.

If the same part of the code shall be tested multiple times,
the recurring parts of the \emph{Inject Arrange} stage may be factored out to the
\emph{Inject Setup} stage.
Finally, if the unit test framework does not automatically clear injected code
after the tests are finished, 
then all code injections need to be manually cleared in the
\emph{Tear down} stage.

The only necessary modification to the system under test is the addition of
comments to the code, indicating where to instrument the code for the injection tests.
Comments are unlikely to break or alter the existing code\footnote{%
An example where comments could break code are old \emph{BASIC} dialects 
which used  hard coded line numbers
to specify the control flow.},
particularly when 
some special style (e.\,g.~\texttt{\mcomment{<!TEST1!>}}) is used, such that changes can be easily tracked
and the comment`s special role is clearly indicated.
If comments were accidentally removed, 
the accompanying test suite would fail, thus
indicating the problem.

\subsection{Consequences}
The refactoring of legacy code becomes easier and less likely to break existing code.

\pagebreak
\subsubsection{Benefits}
\nopagebreak
\begin{itemize}[leftmargin=*]
\item[$+$] Injection testing only poses some requirements on the programming language,
but nearly none on the coding style. 
Thus, the pattern can be applicable in situations where other approaches fail.
This is especially important when refactoring legacy code.
\item[$+$] Interception points can be placed at nearly arbitrary positions in the code,
thus enabling the testing of arbitrary parts of the code.
\item[$+$] Using the injection seam can be faster than the pre-processor or link seam. 
No additional compilation/linking cycles are needed,
but also code sections not relevant to the test can simply be skipped.
\end{itemize}

\subsubsection{Drawbacks}
\begin{itemize}[leftmargin=*]
\item[$-$]
Without sufficient support from the programming language
or the development environment,
injection testing may not be possible or only partly possible,
e.\,g.~ whenever a correspondence the set of key functions cannot be implemented fully.


\item[$-$]
Injection tests are 
more strongly coupled to the system under test than classical unit tests.
In particular, variable names, line numbers, and labels 
may have to be spelled out explicitly in the injection test.
Data prior private to functions (like local variables) become public to the injection test;
In other terminology, the injection test becomes a friend of the system under test.

\item[$-$]
Code injection may degrade the system under test`s performance, 
as $(a)$ the code must be supervised by some means to allow code injection,
and $(b)$ the code can not be compiled with all possible optimizations. 

\item[$-$] Some manual editing of the code is still required, and thus, the risk of unwittingly altering its behaviour is not fully mitigated.

\end{itemize}

\FloatBarrier
\section{Examples}

\label{sec_inject_test}
In this section we give two examples
on using our injection testing pattern
for deriving a unit test and an integration test.
Apart from the functions proposed in section~\ref{sec_solution}, 
we use the following tools:
\begin{itemize}
\item A method to easily save snapshots of some state
\item A unit test framework for automated unit tests
\item An assertion framework to easily write assertions
\end{itemize}


\subsection{Example 1: Unit test}
\label{sec_inject_unit}
We demonstrate the injection testing pattern on the dummy legacy function \texttt{foo}
given in Listing~\ref{lst_foo}, using Matlab style code. 
The task is to refactor parts of the body of \texttt{foo} out into a sub function 
while touching the code as little as possible. 
We only present the relevant lines of code.

\begin{figure}[b!]
\begin{tlisting}{Function \texttt{foo}}{lst_foo}
function foo( a1, a2 )
  % lots of code 
  sum = 0
  for i = 1:a1  <@{\ecomment{\% loop over i = 1 to a1 (both included)}}@>
    sum = sum + i; end
  % lots of code
\end{tlisting}
\Description{Listing of foo}
\end{figure}

\paragraph{Write a succeeding unit test}
The function \texttt{foo} is not suited for classical unit tests, 
since the lines of code to be tested are not accessible via standard means.

In the first step we add the labels
\texttt{\mcomment{<FOO:1>}} and \texttt{\mcomment{<FOO:2>}}
which indicate where to instrument the code
with injection tests, see Listing~\ref{lst_foo_aug}.

\begin{figure}[htb!]
\begin{tlisting}{Function \texttt{foo} augmented}{lst_foo_aug}
function foo( a1, a2 )
  % lots of code 
  % <FOO:1>  <@{\ecomment{\% comment for injection testing}}@>
  sum = 0
  for i = 1:a1
    sum = sum + i; end
  % <FOO:2>  <@{\ecomment{\% comment for injection testing}}@>
  % lots of code
\end{tlisting}
\Description{Listing of foo prepared for injection testing}
\end{figure}

In this form, the code is ready for injection testing
and we can write a unit test suite.
An example unit test suite is given in Listing~\ref{lst_foo_test}.

\begin{figure}[htb!]
\begin{tlisting}{Unit test suite for \texttt{foo}}{lst_foo_test}
% setup
  gotoat( 'foo', 'goto','<FOO:1>' )
  captureat( 'foo', 'at','<FOO:2>', 'var','sum' )    
  returnat( 'foo', 'at','<FOO:2>' )

%% test 1
  % arrange
  assignat( 'foo', 'at','<FOO:1>', 'a1',15 )

  % act
  foo()
  X = captureat()  <@{\ecomment{\% obtain captured values}}@>

  % assert
  EXPECT_EQ( X.FOO2, 120 )  <@{\ecomment{\% compare result}}@>

%% test 2
  % arrange
  assignat( 'foo', 'at','<FOO:1>', 'a1',0 )

  % act
  foo()
  X = captureat()

  % assert
  EXPECT_EQ( X.FOO2, 0 )
    
%% tear down
  clearat( 'foo' )
\end{tlisting}
\Description{Listing of unit tests for foo}
\end{figure}

\paragraph{What happens in Listing~\ref{lst_foo_test}}
In the \texttt{\mcomment{\%\%~setup}} part, 
we collect the code which is shared among both unit tests,
\texttt{\mcomment{\%\%~test~1}} and \texttt{\mcomment{\%\%~test~2}}.
The function \texttt{\kw{gotoat}} injects code such that, after entering
\texttt{foo} the control flow immediately continues at the line with the comment 
\texttt{\mcomment{<FOO:1>}}.
The function \texttt{\kw{captureat}} injects code such that the value of the variable \texttt{sum} 
at the line with comment \texttt{\mcomment{<FOO:2>}} is stored for later retrieval.
The function \texttt{\kw{returnat}} injects code so that the function~\texttt{foo} 
returns to the caller site whenever the control flow reaches line \texttt{\mcomment{<FOO:2>}}.

In the~\texttt{\mcomment{\%\%~test~1}} part we arrange the data to be injected into the function foo;
at the line with comment \texttt{\mcomment{<FOO:1>}}
the variable \texttt{a1} will be assigned the value \texttt{15}.
Afterwards we execute the system under test by calling it
and retrieve the stored data by~\texttt{\kw{captureat}}.
In the \texttt{\mcomment{\%\%~assert}} section we check whether the retrieved value of \texttt{sum} 
equals our expected value \texttt{120}.
The second test~\texttt{\mcomment{\%\%~test~2}} follows the same pattern.

In the \texttt{\mcomment{\%\% tear down}} part, 
after the injection tests have finished,
we clean up using~\texttt{\kw{clearat}}
which removes all injected code from the function~\texttt{foo}.

\paragraph{Refactor}
Having our unit tests in place we can safely refactor the function \texttt{foo},
as given in Listing~\ref{lst_foo_re}.
Afterwards it is usually necessary to also refactor the unit test suite, 
as indicated in Listing~\ref{lst_foo_test_re}.

\begin{figure}[htb!]
\begin{tlisting}{Refactored \texttt{foo}}{lst_foo_re}
function foo( a1, a2 )
  % lots of code 
  % <FOO:1>
  sum = sum0( a1 )
  % <FOO:2>
  % lots of code
  
function x = sum0( x )
  x = x * ( x + 1 ) / 2
\end{tlisting}
\Description{Listing of refactored foo}
\end{figure}

\begin{figure}[htb!]
\begin{tlisting}{Refactored unit tests for \texttt{foo}}{lst_foo_test_re}
%% test 1
  a0 = 15  % arrange
  sum = sum0( a0 )  % act
  EXPECT_EQ( sum, 120 )  % assert

%% test 2
  a0 = 0  % arrange
  sum = sum0( a0 )  % act
  EXPECT_EQ( sum, 0 )  % assert
\end{tlisting}
\Description{Listing of refactored unit tests of foo}
\end{figure}

\FloatBarrier
\subsection{Example 2: Integration test}
\label{sec_inject_integration}

Another example for using the injection test pattern 
is the gradual refactoring of the function \texttt{bar}, 
whose functionality is not apparent to the programmer, 
see Listing~\ref{lst_bar}.

\paragraph{Write a succeeding unit test}

\begin{figure}[htb!]
\begin{tlisting}{Function \texttt{bar}}{lst_bar}
function bar( a )
  % 1000 lines of code
  % <BAR:0>
  % another 1000 lines of code
  % <BAR:1>
\end{tlisting}
\Description{Listing of bar}
\end{figure}

Using injection testing, 
we store the full state of the program at various locations
when run in its initial form, i.e.~before refactoring.
After refactoring we compare the saved state with the new state.
If they coincide, we can assume that the behaviour 
of the system under test did not change.
An exemplary unit test suite is given in~\ref{lst_bar_test}.

\begin{figure}[htb!]
\begin{tlisting}{Integration tests for \texttt{bar}}{lst_bar_test}
%% test
% setup
  captureat( 'bar', 'at','<BAR:0>' )
  captureat( 'bar', 'at','<BAR:1>' )
  
%% test 10
  % arange
  a = 10
  
  % act
  sut( a )
  X = captureat()  <@{\ecomment{\% obtain captured values}}@>

  % assert
  EXPECT_EQ( CACHE('BAR0_10', X.BAR0), X.BAR0 )
  EXPECT_EQ( CACHE('BAR1_10', X.BAR1), X.BAR1 )
  
%% test 20
  % arange
  a = 20
  
  % act
  bar( a )
  X = captureat()  <@{\ecomment{\% obtain captured values}}@>

  % assert
  EXPECT_EQ( CACHE('BAR0_20', X.BAR0), X.BAR0 )
  EXPECT_EQ( CACHE('BAR1_20', X.BAR1), X.BAR1 )
  
%% tear down
  clearat( 'bar' )
\end{tlisting}
\Description{Listing of integration tests for bar}
\end{figure}

\paragraph{What happens in Listing~\ref{lst_bar_test}}
The only substantial difference to our unit test suite for~\texttt{foo}
is the use of the helper function \texttt{\kw{CACHE}}.
This is just a thin convenience wrapper for storing data to disk,
used as follows:
When a file with name equal to its first argument does not exist,
it stores the value of the second argument to disk.
Otherwise, it discards the second argument 
and loads the stored data from disk.

This time we inject code such that the whole workspace of the function~\texttt{bar} is captured
whenever the control flow reaches the lines 
with comments~\texttt{\mcomment{<BAR:0>}} and~\texttt{\mcomment{<BAR:1>}}.
The call \texttt{X = \kw{captureat}()} in the \texttt{\mcomment{\% act}} section 
then retrieves the stored data and stores it in~\texttt{X}.
The assert section now compares the two stored states 
with the snapshot taken from before refactoring.

\FloatBarrier
\section{Remarks about the implementation}
\label{sec_implementation}
Only some programming languages, e.\,g.~\emph{Java}, support code injection directly
When there is no language support,
it is often possible to use the 
debugger 
for implementing code injection. 
Similar approaches are in use for tracing function calls in \emph{Matlab}~\cite{Isa2016}, 
and for implementing Mutation testing in \emph{Java}~\cite{Schu2009}. 
The relevant features of the debugger are:
\begin{itemize}
\item Running a program step by step
\item (Conditionally) stopping the program at a certain point
\item Inspecting the current state
\item Jumping to a certain point 
\end{itemize}

Often, these features can be accessed 
from within the programming language,
e.\,g.~in interpreted languages like \emph{Matlab}, \emph{Python}, or~\emph{R}.
In other languages or IDE`s the debugger can be controlled using third party libraries,
e.\,g.~\emph{libgdb} for \emph{gdb} (discontinued 1993)~\cite{sw_libgdc},
\emph{lldb} for the \emph{LLVM} toolchain~\cite{sw_lldb},
Windows Debugger (\emph{WinDbg}) for \emph{Windows}~\cite{sw_windbg}%
.
Failing that, one can usually still write macros 
in some scripting language for the debugger,
e.\,g.~\emph{GDB\textbackslash MI} for \emph{gdb}~\cite{sw_gdbmi}.

\section{Implementation in Matlab}
\label{sec_ttest}

We implemented the set of key function using Matlab`s debugger and conditional breakpoints.
Matlab`s conditional breakpoints evaluate a string at run-time. 
If the result is \emph{truthy}\footnote{%
A \emph{truthy} value is a value which implicitly evaluates to \texttt{true}, 
for example in an \texttt{if} condition;
e.\,g.~\texttt{true}, \texttt{1} or an array with only non-zero values.
Contrary, a \emph{falsy} value implicitly evaluates to \texttt{false};
e.\,g.~\texttt{false}, \texttt{0}, or an array with at least one zero. 
}, the code run is stopped at that location;
but when it is \emph{falsy}, the code run continues normally.
Using conditional breakpoints for code injection in Matlab
has some restrictions:
\begin{enumerate}[leftmargin=*]
\item\label{inject_handle}
  Conditional breakpoints only accept valid Matlab commands, but not arbitrary anonymous functions. 

  
\item\label{inject_falsy}
  The injected code must always return a \emph{falsy} value 
  in order to avoid the debugger stopping its execution.

\item\label{inject:error}
  If the injected code throws an error, it is caught automatically by
  Matlab and the program run stops, i.e.~the debugger starts.
  
\item\label{inject:before}
  Injected code is always executed before the code at the injected line.
  Code cannot be injected between statements.

\end{enumerate}

To execute anonymous functions, 
we store them in a persistent variable in some function, 
and generate a string which then executes that anonymous function.
To ensure that the return value of the injected code 
is
\texttt{false}, 
it gets wrapped in a function returning \texttt{false} and evaluated by
\texttt{evalin(~\qstring{caller},~\_\_~)}\footnote{
The function~\texttt{evalin(~\qstring{caller},~cmd~)} 
executes a command \texttt{cmd} in the callers workspace,
and in particular has access to the callers workspace.}.  
Errors thrown by the injected code are caught inside the function which evaluates
the string or anonymous function.



\subsection{%
\texorpdfstring{Example implementation of \texttt{\kw{evalat}}}%
{Example implementation of evalat}%
}
To illustrate the execution of anonymous functions in Matlab, 
and thus 
execution of
arbitrary code,
a minimum implementation of a function \texttt{\kw{evalat}} is given in Listing~\ref{lst_evalat}. 
The key functions~\texttt{\kw{captureat}}, \texttt{\kw{assignat}} 
both can be derived from this one.
Note, to avoid parsing the inputs,
the interface of \texttt{\kw{evalat}} is different
from the interface of the~\texttt{\kw{...at}} functions
in the listings above.

\begin{figure}[htb!]
\begin{tlisting}{\texttt{\kw{evalat}}}{lst_evalat}
function ret = evalat( fun, lne, h );
  persistent <@{\color{persistentcolor}cache}@>;
  if( nargin==0 );
    ret = <@{\color{persistentcolor}cache}@>;
    return; end;    
  <@{\color{persistentcolor}cache}@> = h;
  h = ['returnfalse( ' ...
       '  assign(''ttest_handle'',evalat()) ) ||'...
       'returnfalse( ttest_handle() ); '];
  dbstop( 'in',fun, 'at',num2str(lne), 'if',h );

function ret = returnfalse( varargin );
  ret = false;
  
function ret = assign( nme, val );
  try; ret = evalin( 'caller', [name ';'] ); 
  catch; ret = []; end;
  assignin( 'caller', name, value );
\end{tlisting}
\Description{Listing of evalat}
\end{figure}

\paragraph{What happens in Listing~\ref{lst_evalat}}
The function \texttt{\kw{evalat}} accepts three arguments, 
\texttt{fun} is the function where code shall be injected,
\texttt{lne} is the line number where code shall be injected,
\texttt{h} is an anonymous function 
to be executed at the specified position.

Upon calling with three arguments, the anonymous function~\texttt{h} is stored in the 
persistent variable~\texttt{\color{persistentcolor}cache}.
A persistent variable retains its values between function calls.
Then, the function \texttt{dbstop} adds in the function~\texttt{fun}
at the specified position~\texttt{lne}
a conditional breakpoint, 
which will execute the code listed in Listing~\ref{lst_evalat_1};
We put those lines of code in its own listing for better readability.

\begin{figure}[htb!]
\begin{tlisting}{Code of conditional breakpoint in \texttt{\kw{evalat}}}{lst_evalat_1}
returnfalse( assign('ttest_handle',evalat()) ) ||...
returnfalse( ttest_handle() );
\end{tlisting}
\Description{Listing of the code of the conditional breakpoint in evalat}
\end{figure}

Now, when the function~\texttt{fun} is called 
and the program flow reaches line~\texttt{lne}
the function \texttt{\kw{evalat}} is called without arguments.
Thus, \texttt{evalat} returns the value 
of the persistent variable \texttt{\color{persistentcolor}cache},
this is exactly the anonymous function~\texttt{h} we want to execute.
The anonymous function~\texttt{h} is passed to the function \texttt{\kw{assign}}, 
which creates a variable with name~\texttt{\mstring{'ttest\_handle'}} 
in the workspace of the function~\texttt{fun}.
All of this code is wrapped inside a call to~\texttt{\kw{returnfalse}},
which ensures that always \texttt{false} is returned.

A conditional breakpoint only stops when the injected code returns \texttt{true}. 
Since the first part of the injected code returns \texttt{false}
the second part is evaluated.
Now the just assigned variable~\texttt{ttest\_\-handle}, 
which is our anonymous function~\texttt{h}, is executed.
The result of the anonymous function~\texttt{h}
is passed again to~\texttt{\kw{returnfalse}}
which again ensures that \texttt{false} is returned.
Thus, the debugger does not stop the program.

A usage example of~\texttt{\kw{evalat}} is given in Listing~\ref{lst_evalat_usage}.

\begin{figure}[htb]
\begin{tlisting}{Usage example of \texttt{\kw{evalat}}}{lst_evalat_usage}
>> evalat( 'surf', 1, @() disp('Hello World!') );
>> surf( membrane );
<@\mstring{Hello World!}@>  <@{\ecomment{\% and the Matlab membrane is plotted}}@>
\end{tlisting}
\Description{Listing of usage example of evalat}
\vspace*{-4mm}
\end{figure}

\subsection{%
\texorpdfstring{Example implementation of \texttt{\kw{returnat}}}%
{Example implementation of returnat}%
}
A more involved, but shorter example shows
how to programmatically return early from a function.
The idea is to provoke an error at a user defined position in a function and catch the exception.
As already noted, simply throwing an error in some injected code would not work,
since, whenever injected code throws, the debugger stops the program.
Instead, we have to make sure that an error is thrown
after the injected code was executed.
This we achieve by some tough means; we clear the function`s workspace, 
see Listing~\ref{lst_returnat}.%

\begin{figure}[htb!]
\begin{tlisting}{\texttt{\kw{returnat}}}{lst_returnat}
function ret = returnat( fun, lne );
  if( nargin==0 );
    evalin( 'caller', 'clear' );
    ret = false;
  else;
    dbstop( 'in',fun, 'at',num2str(lne), ...
            'if','returnat' ); 
    try; eval( fun );
    catch me; disp( me ); end; end;
\end{tlisting}
\Description{Listing of returnat}
\end{figure}

\paragraph{What happens in Listing~\ref{lst_returnat}}
The function \texttt{\kw{returnat}} accepts two arguments,
\texttt{fun} is the function which shall be executed,
\texttt{lne} is the line number at which we want to return.
When called with two arguments,
\texttt{dbstop} adds a conditional breakpoint in the function~\texttt{fun}
at the specified position~\texttt{lne},
which executes a call to~\texttt{\kw{returnat}} without arguments.
Afterwards the function~\texttt{fun} is called.
When program flow reaches the specified location~\texttt{lne},
\texttt{\kw{returnat}} is called without arguments.
Thus, the workspace of \texttt{fun} is cleared by
executing~\texttt{evalin( \mstring{'caller'}, \mstring{'clear'} )}. 
Finally \texttt{false} is returned, so that the debugger does not stop the program.
The next time a variable is accessed in the function~\texttt{fun},
an error is thrown, therefore the~\texttt{fun} returns.
This error is caught in the~\texttt{catch} block 
in the function~\texttt{\kw{returnat}}.
In our example implementation we display the caught error, 
but any other code is equally possible.
A usage example
is given in Listing~\ref{lst_returnat_usage}.

\begin{figure}[htb!]
\begin{tlisting}{Usage example of \texttt{\kw{returnat}}}{lst_returnat_usage}
>> returnat( 'spy', 42 );
  MException with properties:
  identifier: 'MATLAB:refClearedVar'
     message: 'Reference to a cleared variable.'
        file: 'spy.m::42'
\end{tlisting}
\Description{Listing of usage example of returnat}
\end{figure}

\subsection{\emph{TTEST}}

Functionality for implementing the code injection pattern, specifically the key functions
\texttt{\kw{assignat}}, \texttt{\kw{captureat}}, \texttt{\kw{evalat}}, 
\texttt{\kw{returnat}}, are contained in the unit test framework \emph{TTEST} for Matlab and Octave.
\emph{TTEST} has been written specifically with testing code in a scientific context in mind. 
It supports the testing of scripts, local and sub functions,
has utilities for caching results for integration tests, and
adds support for injection testing
and partly for design by contract~\cite{Meyer1992}.



\emph{TTEST} is published under a permissive open source license and available 
at~\href{https://gitlab.com/tommsch/TTEST}%
{\emph{gitlab.com/\-tommsch/\-TTEST}}. The full documentation of \emph{TTEST}, 
together with a comparison of Matlab unit test frameworks,
can be found in~\cite{fw_ttest}.

The following projects use \emph{TTEST} (list non exhaustive):
\begin{itemize*}[leftmargin=*]
\item \emph{Auditory Modelling Toolbox} (Ver.~$1.1$)~\cite{usage_amt}
\item \emph{ttoolboxes}~\cite{usage_tjsr}
\item \emph{Large Time Frequency Analysis Toolbox} (Ver.~$> 2.4$)~\cite{usage_ltfat}
\end{itemize*}

\section{Conclusion}
We presented a pattern for the injection-based refactoring as a means for 
handling otherwise not testable code, 
along with a set of functions suitable for its implementation. 
We gave examples on its usage and practical advice for their implementation 
in scripting languages via making use of the debugger`s functionalities. 
We provide a free implementation of the pattern`s key functionality 
in our \emph{TTEST} unit testing framework.

Further work comprises enhancements of the usability and customizability 
of the pattern by improving on the underlying functionality in the toolbox, 
e.\,g.~by implementing a \texttt{\kw{gotoat}} function, 
allowing for the direct execution of arbitrary sections in the code. 

\begin{acks}
This work has been supported by
the Austrian Science Foundation (FWF) 
grant P33352-N,
and by the European Union (EU) within the project SONICOM 
grant 101017743, RIA action of Horizon 2020.
\end{acks}

\balance


\newcommand{\doi}[1]{\href{https://doi.org/#1}{doi:~#1}}
\newcommand{\doitwo}[2]{\href{https://doi.org/#1}{doi:~#2}}
\newcommand{\arxiv}[1]{\href{https://arxiv.org/abs/#1}{arXiv:~#1}}

\makeatletter
\newcommand*{\textoverline}[1]{$\overline{\hbox{#1}}\m@th$}
\makeatother


\end{document}